\documentstyle[12pt]{article}
\newcommand{\RR}{$\rm l\!R\;\;$}

\newcommand{\eut}{
\begin{picture}(3,3)(-2,-2)
\put(-2,-15){$\tilde{}$}
\put(-5,-2){$\eta$}
\end{picture}}

\newcommand{\fu}{
\begin{picture}(9,5)(0,0)
\put(2.5,10){\tiny 1}
\put(0,0){F}
\end{picture}}

\newcommand{\fd}{
\begin{picture}(9,5)(0,0)
\put(2.5,10){\tiny 2}
\put(0,0){F}
\end{picture}}

\newcommand{\bu}{
\begin{picture}(9,5)(0,0)
\put(2.5,12){\tiny 1}
\put(0,0){$\tilde{{\rm B}}$}
\end{picture}}

\newcommand{\bd}{
\begin{picture}(9,5)(0,0)
\put(2.5,12){\tiny 2}
\put(0,0){$\tilde{{\rm B}}$}
\end{picture}}

\newcommand{\au}{
\begin{picture}(9,5)(0,0)
\put(3,10){\tiny 1}
\put(0,0){A}
\end{picture}}

\newcommand{\ad}{
\begin{picture}(9,5)(0,0)
\put(2.5,10){\tiny 2}
\put(0,0){A}
\end{picture}}

\newcommand{\cu}{
\begin{picture}(11,5)(0,0)
\thicklines
\put(0,9){\line(1,-2){5}}
\put(0,9){\line(1,0){10}}
\thinlines
\put(5,-1){\line(1,2){5}}
\put(3.7,3.7){\tiny 1}
\end{picture}}

\newcommand{\cd}{
\begin{picture}(11,5)(0,0)
\thicklines
\put(0,9){\line(1,-2){5}}
\put(0,9){\line(1,0){10}}
\thinlines
\put(5,-1){\line(1,2){5}}
\put(3.7,3.7){\tiny 2}
\end{picture}}

\newcommand{\piu}{
\begin{picture}(9,5)(0,0)
\put(2,9){\tiny 1}
\put(0,0){$\tilde{\pi}$}
\end{picture}}

\newcommand{\pid}{
\begin{picture}(9,5)(0,0)
\put(2,9){\tiny 2}
\put(0,0){$\tilde{\pi}$}
\end{picture}}

\begin{document}

\baselineskip=22pt plus 0.2pt minus 0.2pt
\lineskip=22pt plus 0.2pt minus 0.2pt

\hspace{4.5in}CGPG-93/9-5
\vspace{1in}

\begin{center}
 \Large
General Relativity as a Theory of Two Connections\\

\vspace*{0.35in}

\large

J. Fernando Barbero G.\footnote{Also at Instituto de Matem\'aticas y
F\'{\i}sica Fundamental, C.S.I.C. Serrano 119--123, 28006 Madrid, Spain}

\vspace*{0.25in}

\normalsize

104 Davey Lab., Penn State University, \\
University Park, P.A. 16802, U.S.A. \\

\vspace{.5in}
September 28, 1993\\
\vspace{2in}
ABSTRACT

\end{center}

We show in this paper that it is possible to formulate General Relativity in a
phase space coordinatized by two $SO(3)$ connections. We analyze first the
Husain-Kucha\v{r} model and find a two connection description for it.
Introducing a suitable scalar constraint in this phase space we get a
Hamiltonian formulation of gravity that is close to the Ashtekar one,
from which it is derived, but has some interesting features of its own. Among
them a possible mechanism for dealing with the degenerate metrics and a
neat way of writing the constraints of General Relativity.

\vspace{2cm}
\pagebreak

\setcounter{page}{1}

\section{Introduction}

After the introduction by Ashtekar \cite{Ash} of a new set of variables to
describe the gravitational field we have a picture of classical and quantum
gravity that is very different from the geometrodynamical point of view that
has prevailed during the last decades (see, for example, \cite{MTW}
\cite{ADM}). The possibility of
restating most of the problems faced by the old program in the new language and
the insights gained because of the geometrical nature of the Ashtekar variables
have shed light on many important issues and have helped to gain a new
perspective in space-time Physics. A key ingredient of the Ashtekar
approach is
the use of a connection as the basic variable to describe the
gravitational field (the frame
fields appear only as the momenta canonically conjugate to this connection).

{}From a geometrical point of view, connections and metrics are very different
objects. They find also very different uses in Theoretical Physics. Whereas
metrics appear mainly in the context of gravity, connections are a key
element in the description of the electroweak and strong interactions. There
are lots of results about connections that have no analogue in the case of
metrics and can be successfully exploited in the Ashtekar formalism.
For example, Wilson loops, that were introduced in the context of
Yang-Mills theories, are at the root of the loop-variables approach to the
quantization of gravity. The fact that gravity can be described in terms
of connections is very appealing, as emphasized by Ashtekar
\cite{Ashlibro}, because for the first time all four interactions have
something
in common in their mathematical formulations; all of them can be formulated in
a
Yang-Mills phase space.

The purpose of this paper is to show another description of gravity that,
although close to the Ashtekar formulation, from which it is derived, has some
features of
its own that may make it useful in order to better understanding gravity and
generally covariant theories.
For example, the issue of the degenerate metrics can be related, within the
framework we are going to work, to the
non-degeneracy of the symplectic 2-form that appears in the
Hamiltonian description of the theory. An important point in our approach is
the fact that
the phase space is now different from the usual one (connection-densitized
triad). In fact, the phase space variables will be now two $SO(3)$ connections.
As a consequence of
this, the symplectic structure will be non-trivial (notice that connections
alone cannot be
momenta because they have zero density weight). The ultimate goal of this
approach is that of taking advantage of the availability of geometrical objects
absent (or, at least, not obvious) in the Ashtekar formulation that might
eventually allow us to find
a set of elementary variables suitable for the quantization of the theory.
This idea is in line with the suggestions
made by Isham \cite{Ish} that non-canonical algebras may be the way to quantize
gravity and is close, in this sense, to the loop-variables approach.
We want to enphasize from the beginning that the final formulation of gravity
that we give in the paper is purely in terms
of connections but different from the Capovilla-Dell-Jacobson one
\cite{CDJ}. The starting point of these authors is a pure connection
action but the Hamiltonian formulation that they find essentially coincides
with the Ashtekar one because they use the same variables in the phase space.

The paper is structured as follows. In section 2 we show that the
Husain-Kucha\v{r} model can be interpreted as a theory of two connections.
This is an interesting toy model for the
study of gravity and diffeomorphism invariant theories in general and it is
simple to analyze because it has no dynamics. All the complications
associated with the Hamiltonian constraint are absent simply
because there is no scalar constraint in the theory. We will give a new action
for this model written in terms of two connections and study the
Hamiltonian formulation, phase space structure and constraints of the theory.
The action that we will use bears some resemblance with the BF-actions studied
by Horowitz \cite{Hor} and other authors as exactly solvable diff-invariant
theories.

In section 3 we show that there is a change of coordinates that will allow us
to
recover the usual Gauss law and vector constraint written in terms of the usual
Ashtekar variables. This transformation will be used to write the Hamiltonian
constrain in terms of the two connections. In passing we will find
an interesting type of canonical transformation in the new phase space.

Section 4 is devoted to the study of gravity in the two-connection
formulation. An important issue that will be considered here is that of the
degenerate metrics. The key observation is that the
necessary and sufficient condition for the 3-metric to be non-degenerate is the
non-degeneracy of the symplectic 2-form. It is this consistency condition for
the Hamiltonian formulation that forces us to restrict ourselves to
non-degenerate metrics. Taking advantage of this non-degeneracy we will find
an appealing way of writing the constraints of general relativity.

We end the paper with the conclusions and a brief discussion of the open
questions related to the two connection approach to gravity.

\section{The Husain-Kucha\v{r} Model as a Two Connection Theory}

The Husain-Kucha\v{r} model \cite{Huku} is a very interesting example of a
diffeomorphism invariant theory. At variance with most diff-invariant
theories appearing in the literature it has an infinite number of local degrees
of freedom (three per space point) and yet it is simpler than general
relativity because the scalar constraint (one of the key sources of
trouble in gravity) is not present. The constraints of the model are the Gauss
law, that generates $SO(3)$ gauge transformations and the vector constraint
that
(combined with the Gauss law) generates diffeomorphisms on the "spatial" slices
of the 3+1 decomposition. These
constraints are first class in Dirac's terminology \cite{Dirac}. The absence of
the Hamiltonian constraint can be interpreted by saying there is no dynamics,
or
rather, time evolution. In this section we show that the Husain-Kucha\v{r}
model
can be described as a pure connection theory. This is somewhat similar to what
Capovilla, Dell, and Jacobson did for gravity \cite{CDJ}, the key
difference is
that now we will introduce {\it two} connections instead of one.

The Husain-Kucha\v{r} action is:

\begin{equation}
S=\frac{1}{2}\int_{\cal M}d^{4}x\;\tilde{\eta}^{abcd}F_{ab}^{i}e_{c}^{j}
e_{d}^{k}\epsilon_{ijk}
\label{1}
\end{equation}

\noindent where $e_{a}^{i}$ are $SO(3)$ valued frame fields, $F_{ab}^{i}$ is
the
curvature of the $SO(3)$ connection $A_{a}^{i}$ (given by $F_{ab}^{i}=
2\partial_{[a}A_{b]}^{i}+\epsilon^{ijk}A_{aj}A_{bk})$, $\tilde{\eta}^{abcd}$
is the
4-dimensional Levi-Civita tensor density (we will use Ashtekar's notation and
represent the density weights with tildes above or below the fields) and
$\epsilon^{ijk}$ is the internal Levi-Civita tensor. In the following we will
restrict our attention to space-time 4-manifolds of the form ${\cal M}=
\Sigma\times$\RR with $\Sigma$ a compact 3-manifold. Tangent space indices
will be represented by Latin letters from the beginning of the alphabet and
internal indices with Latin letters from the middle of the alphabet. We will
make no distinction between the indices of 4-dimensional and 3-dimensional
objects; it will be clear from the context which kind of field we are
talking about.

The key observation to pass from (\ref{1}) to the two-connection action is
that a $SO(3)$ connection $A_{a}^{i}$ has the same index structure as the
frame fields $e_{a}^{i}$ (remember that connections, in the adjoint
representation, carry always two indices.
It is only the fact that $SU(2)$ connections are antisymmetric in their
internal indices and the availability of $\epsilon_{ijk}$ that allows us to
represent both indices as one). One can then conceivably write $e_{a}^{i}$ as
the difference of two $SO(3)$ connections $\au_{a}^{i}$, $\ad_{a}^{i}$.
If one further imposes the condition that the theory should be
symmetric under the interchange of both connections it is natural to consider
the action:

\begin{equation}
S=\frac{1}{2}\int_{\cal M}d^{4}x\;\tilde{\eta}^{abcd}\fu_{ab}^{i}
\fd_{cd\;i}
\label{2}
\end{equation}

\noindent where $\fu^{i}_{ab}$, $\fd^{i}_{ab}$ are the curvatures of
$\au^{i}_{a}$ and $\ad^{i}_{a}$ respectively. In order to see that (\ref{2})
and (\ref{1}) describe the same theory we write:

\begin{equation}
\fd^{i}_{ab}=\fu^{i}_{ab}+ 2\;\cu_{[a}e_{b]}^{i}+\epsilon^{ijk}e_{aj}e_{bk}
\label{2b}
\end{equation}

\noindent where $e_{ai}=\ad_{ai}-\au_{ai}$ and $\cu_{a}$ is the covariant
derivative built with the connection $\au_{a}^{i}$ that acts only on internal
indices as\footnote {We may extend the action of $\cd_{a}$ to space-time
indices by introducing a torsion free connection $\Gamma^{c}_{ab}$. Our results
will be independent of this extension.}:
$\cu_{a}\lambda_{i}=\partial_{a}\lambda_{i}+\epsilon_{i}^{\;\;jk}\au_{aj}
\lambda_{k}$ (we define $\cd$ in an analogous way).

\noindent Introducing (\ref{2b}) in (\ref{2}) we get:

\begin{equation}
S=\frac{1}{2}\int_{\cal M}d^{4}x\;\tilde{\eta}^{abcd}\left[\fu_{ab}^{i}
\fu_{cdi}+2\fu_{ab}^{i}\cu_{c}e_{di}+\fu_{ab}^{i}e_{c}^{j}e_{d}^{k}
\epsilon_{ijk}\right]
\label{3}
\end{equation}

\noindent The first term in the previous expression is topological and thus we
can discard it when checking that (\ref{1}) and (\ref{2}) give the same
dynamics. The covariant derivative in the second term can be integrated by
parts and then the Bianchi identity $\cu_{[a}\fu_{bc]}^{i}=0$ tells us that
this term does not contribute to the action either.
Finally the third term
coincides with the one appearing in (\ref{1}) provided that we make the
identifications ${\rm A}_{a}^{i}\equiv\au_{a}^{i}$ and $e_{ai}\equiv\ad_{ai}-
\au_{ai}$.
Notice that in order to get the field equations it is equivalent to vary the
action with respect to ${\rm A}_{a}^{i}$, $e_{a}^{i}$ or $\au_{a}^{i}$,
$\ad_{a}^{i}$.

It is interesting to point out that the action (\ref{2})
becomes a topological invariant when $\au_{a}^{i}=\ad_{a}^{i}$. The field
equations derived from (\ref{2}) take a very simple form:

\begin{eqnarray}
& & \cu_{[a}\fd_{bc]}^{i}=0\nonumber\\
& & \cd_{[a}\fu_{bc]}^{i}=0\label{4}
\end{eqnarray}

\noindent when $\au_{a}^{i}=\ad_{a}^{i}$ they reduce to the Bianchi identities
and hence they are identically satisfied (a reflection of the "triviality" of
the action when $\au_{a}^{i}=\ad_{a}^{i}$). The simple structure of (\ref{4})
suggests some interesting classes of solutions. For example, if
$\au_{a}^{i}$ and $\ad_{a}^{i}$
are such that $\fu_{ab}^{i}=\kappa \fd_{ab}^{i}$ with $\kappa$ constant then
(\ref{4}) is trivially satisfied as a consequence of the Bianchi identities.
This type of solutions is not obvious if one looks at the field equations as
derived from (\ref{1}):

\begin{eqnarray}
& & \tilde{\eta}^{abcd}\epsilon_{i}^{\;\;jk}(\nabla_{a} e_{bj})e_{ck}=0
\nonumber\\
& & \tilde{\eta}^{abcd}\epsilon_{i}^{\;\;jk}e_{aj}{\rm F}_{bck}=0\label{5}
\end{eqnarray}

\noindent (Their equivalence with (\ref{4}) can be easily checked by using
(\ref{2b}), the definition of the curvature as the commutator of covariant
derivatives and the Bianchi identities).\footnote{In (\ref{5}), ${\rm
F}_{ab}^{i}$ is the curvature of the connection ${\rm A}_{a}^{i}$ used to
define
the covariant derivative $\nabla$}.

We will concentrate now on the discussion of the Hamiltonian form of the theory
described by the action (\ref{2}). Although the final description we will get
is equivalent to the usual Husain-Kucha\v{r} model the
phase space will be different now (because instead of the connection and
densitized triads the coordinates in the new phase space will be two $SO(3)$
connections). This may have some interesting consequences when trying to
quantize the theory because we may have now the possibility of finding sets of
elementary variables different from the connection-densitized triad pair or
the loop variables.

In order to perform the 3+1 decomposition we introduce a foliation of $\cal M$
given by 3-surfaces of constant $t$ (where $t$ is a scalar function defined on
$\cal M$). We need also a vector field $t^{a}$ satisfying the condition
$t^{a}\partial_{a}t=1$. With the aid of $t^{a}$ we can write
$\tilde{\eta}^{abcd}d^{4}x=4 t^{[a}\tilde{\eta}^{bcd]}d^{3}x\;dt$ and then
(\ref{2}) becomes:

\begin{equation}
S=\int dt\int_{\Sigma}d^{3}x\left[t^{a}\tilde{\eta}^{bcd}\fu_{ab}^{i}
\fd_{cdi}+t^{c}\tilde{\eta}^{dab}\fu_{ab}^{i}\fd_{cdi}\right]
\label{6}
\end{equation}

\noindent Using now
the identity\footnote{${\cal L}_{t}$ denotes the Lie derivative along the
direction of the vector field $t^{a}$ and $A_{0}^{i}\equiv t^{a}A_{a}^{i}$}
$t^{a}F_{ab}^{i}={\cal L}_{t} A_{b}^{i}-\nabla_{b}
A_{0}^{i}$ we can rewrite (\ref{6}) in as:

\begin{eqnarray}
& & S=\int dt\int_{\Sigma}d^{3}x\left[({\cal L}_{t}\au_{ai})\tilde{\eta}^{abc}
\fd_{bc}^{i}+({\cal L}_{t}\ad_{ai})\tilde{\eta}^{abc}
\fu_{bc}^{i}+\right.\nonumber\\
& & \left.\hspace{6cm}+\au_{0}^{i}\cu_{a}(\tilde{\eta}^{abc}\fd_{bc\;i})+
\ad_{0}^{i}\cd_{a}(\tilde{\eta}^{abc}\fu_{bc\;i})\right]
\label{6b}
\end{eqnarray}

\noindent Where $\au_{0}^{i}\equiv t^{a}\au_{a}^{i}$ and
$\ad_{0}^{i}\equiv t^{a}\ad_{a}^{i}$. The
connections and the curvatures appearing now in (\ref{6}),
(\ref{6b}) are the pull-backs to the 3-surfaces in the 3+1 decomposition of the
4-dimensional objects and $\tilde{\eta}^{abc}$ is the 3-dimensional Levi-Civita
tensor density. From (\ref{6b}) we see that the momenta canonically conjugate
to $\au_{0}^{i}$, $\ad_{0}^{i}$, $\au_{a}^{i}$, and $\ad_{a}^{i}$ are:
\begin{eqnarray}
&\piu_{i}^{\scriptscriptstyle {0}}\equiv
\begin{picture}(40,30)(0,0)
\put(14,7){$\delta L$}
\put(0,-12){$\delta\left({\cal L}_{t} \au_{\scriptscriptstyle {0}}^{i}
\right)$}
\put(0,3){\line(1,0){40}}
\end{picture}
=0\hspace{3cm}
&\piu_{i}^{\scriptscriptstyle {a}}\equiv
\begin{picture}(40,26)(0,0)
\put(14,7){$\delta L$}
\put(0,-12){$\delta\left({\cal L}_{t} \au_{\scriptscriptstyle {a}}^{i}
\right)$}
\put(0,3){\line(1,0){40}}
\end{picture}
=\tilde{\eta}^{abc}\fd_{bc\;i}\equiv\bd_{i}^{a}\nonumber\\
&\pid_{i}^{\scriptscriptstyle {0}}\equiv
\begin{picture}(40,30)(0,0)
\put(14,7){$\delta L$}
\put(0,-12){$\delta\left({\cal L}_{t} \ad_{\scriptscriptstyle {0}}^{i}
\right)$}
\put(0,3){\line(1,0){40}}
\end{picture}
=0\hspace{3cm}
&\pid_{i}^{\scriptscriptstyle {a}}\equiv
\begin{picture}(40,26)(0,0)
\put(14,7){$\delta L$}
\put(0,-12){$\delta\left({\cal L}_{t} \ad_{\scriptscriptstyle {a}}^{i}
\right)$}
\put(0,3){\line(1,0){40}}
\end{picture}
=\tilde{\eta}^{abc}\fu_{bc\;i}\equiv\bu_{i}^{a}\label{7}
\end{eqnarray}

\noindent Where $L$ denotes the Lagrangian. The non-zero canonical Poisson
brackets are:

\begin{eqnarray}
& & \left\{\au_{0}^{i}(x), \piu^{0}_{j}(y)\right\}=
\left\{\ad_{0}^{i}(x), \pid^{0}_{j}(y)\right\}=\delta_{j}^{\;\;i}\delta^{3}
(x,y)\nonumber\\
& &
\left\{\au_{a}^{i}(x), \piu^{b}_{j}(y)\right\}=
\left\{\ad_{a}^{i}(x), \pid^{b}_{j}(y)\right\}=
\delta_{j}^{\;\;i}\delta_{a}^{\;\;b}\delta^{3}(x,y)
\label{8}
\end{eqnarray}

{}From the definition of the momenta given by (\ref{7}) we have the following
primary constraints:

\begin{eqnarray}
&\piu^{0}_{i}=0 \hspace{4cm}& \piu^{a}_{i}=\bd^{a}_{i}\nonumber\\
& \pid^{0}_{i}=0 \hspace{4cm}& \pid^{a}_{i}=\bu^{a}_{i}\label{9}
\end{eqnarray}

\noindent and the Hamiltonian:

\begin{equation}
H=-\int_{\Sigma}d^{3}x\left[\au_{0}^{i}\cu_{a}\bd^{a}_{i}+
\ad_{0}^{i}\cd_{a}\bu^{a}_{i}\right]
\label{10}
\end{equation}

\noindent Following Dirac (see, for example \cite{Dirac}) we introduce a
total Hamiltonian:

\begin{equation}
H_{T}=H+\int_{\Sigma}d^{3}x\left[u^{i}\piu_{i}^{\scriptscriptstyle {0}}+v^{i}
\pid_{i}^{\scriptscriptstyle {0}}+
u_{a}^{i}(\piu^{a}_{i}-\bd^{a}_{i})+v_{a}^{i}(\pid^{a}_{i}-\bu^{a}_{i})\right]
\label{10b}
\end{equation}

\noindent ($u^{i}$, $v^{i}$, $u_{a}^{i}$, and $v_{a}^{i}$ are Lagrange
multipliers) and impose the conservation in time of the constraints (\ref{9})
under the dynamics defined by $H_{T}$. We find in this way the following
secondary constraints:

\begin{eqnarray}
& & \cu_{\!a}\bd^{a}_{i}=0\nonumber\\
& & \cd_{\!a}\bu^{a}_{i}=0\label{10c}
\end{eqnarray}

Finally we must solve the second class constraints of the theory in order to
get
the Dirac brackets. These Dirac brackets will provide us with the symplectic
structure in the final phase space. The outcome of this analysis is the
following. The phase space is spanned by the 3-dimensional $SO(3)$ connections
$\au_{a}^{i}$ and $\ad_{a}^{i}$, the constraint manifold is defined by the
conditions (\ref{10c}) and the symplectic 2-form is:

\begin{equation}
\Omega=2\int_{\Sigma}d^{3}x\;\tilde{\eta}^{abc}\epsilon_{ijk}\left[\ad_{a}^{i}
(x)-\au_{a}^{i}(x)\right]d\au_{b}^{j}(x)\wedge d\ad_{c}^{k}(x)
\label{11}
\end{equation}

\noindent $\Omega$ is obviously closed; however, it may be degenerate if the
determinant of the $9\times 9$ matrix
$\omega_{(a_{\scriptstyle{i}})(b_{\scriptstyle{j}})}=2\;\tilde{\eta}^{abc}
\epsilon_{ijk}\left[\ad_{c}^{k}(x)-\au_{c}^{k}(x)\right]$ is zero. We will
discuss this issue in section 4. Notice that we have 18 components of
$\au_{a}^{i}$
and $\ad_{a}^{i}$ and six first class constraints per space point (this will be
shown bellow) so that we have $\frac{1}{2}(18-6\times2)=3$ degrees of freedom
per point. It is interesting to point out, also, that both the symplectic
structure (\ref{11}) and the constraint hypersurface defined by (\ref{10c}) are
insensitive to the interchange of the connections. This fact proves to be
useful in order to find the constraint functionals that generate the internal
$SO(3)$ transformations and the diffeomorphisms on $\Sigma$. Before discussing
this point we give several identities that the Dirac brackets
satisfy\footnote{We define the Poisson (Dirac) brackets of two phase space
functions f and g as $\{f,g\}=\Omega^{\alpha\beta}\partial_{\alpha}
f\partial_{\beta}g$ where
$\Omega_{\alpha\beta}$ are the components of the symplectic 2-form in some
coordinate system in the phase space, and
$\Omega^{\alpha\beta}\Omega_{\beta\gamma}=-\delta^{\alpha}_{\;\;\gamma}$.} .
They will help in simplifying the computations that follow. The first of them
(which can be directly read from (\ref{11})) are:

\begin{eqnarray}
& 2\epsilon_{ijk}\tilde{\eta}^{abc}\left[\ad_{a}^{i}(x)-\au_{a}^{i}(x)\right]
\left\{\au_{b}^{j}(x),
\ad_{d}^{l}(y)\right\}=\delta_{c}^{\;\;d}\delta_{k}^{\;\;l}\delta^{3}(x,y) &
\nonumber\\
& \left\{\au_{a}^{i}(x), \au_{b}^{j}(y)\right\}=0 & \label{13}\\
& \left\{\ad_{a}^{i}(x), \ad_{b}^{j}(y)\right\}=0 & \nonumber\\
& \left\{\au_{a}^{i}(x), \ad_{b}^{j}(y)\right\}+\left\{\ad_{a}^{i}(x),
\au_{b}^{j}(y)\right\}=0 & \nonumber
\end{eqnarray}

\noindent The first of the previous expressions can be written also as:

\begin{equation}
\left\{\bu^{a}_{i}(x), \ad_{b}^{j}(y)\right\}+
\left\{\bd^{a}_{i}(x),
\au_{b}^{j}(y)\right\}=-\delta_{b}^{\;\;a}\delta_{i}^{\;\;j}\delta^{3}(x,y)
\label{13b}
\end{equation}

\noindent Another two useful expressions are:

\begin{eqnarray}
& & \epsilon_{j}^{\;\;kl}\tilde{\eta}^{bde}\left(\ad_{dk}-\au_{dk}\right)
\left\{\bd^{a}_{i}(x),\au_{el}(y)\right\}=\nonumber\\
& &
\hspace{4.5cm}=\tilde{\eta}^{abc}\delta_{ij}\;\;
\partial^{\!\!\!\!\!\!\!\!x}_{c}\delta^{3}(x,y)-\tilde{\eta}^{abc}
\epsilon_{ij}^{\;\;\;\;k}\ad_{ck}(x)\delta^{3}(x,y)\label{14}\\
& & \left\{\bu^{a}_{i}(x), \bd^{b}_{j}(y)\right\}+
\left\{\bd^{a}_{i}(x), \bu^{b}_{j}(y)\right\}=0\nonumber
\end{eqnarray}

The generating functionals of the internal gauge transformations and
diffeomorphisms are:

\begin{eqnarray}
& & G(N)=-\int_{\Sigma}d^{3}x
N^{i}\left[\cu_{b}\bd^{b}_{i}+\cd_{b}\bu^{b}_{i}\right]\nonumber\\
& & D(\vec{N})=\int_{\Sigma}d^{3}x N^{a}\left[\ad_{a}^{i}\cd_{b}\bu^{b}_{i}+
\au_{a}^{i}\cu_{b}\bd^{b}_{i}\right]\label{15}
\end{eqnarray}

\noindent The first of these expressions is (modulo a numerical factor) the
simplest linear combination of
(\ref{10c}) symmetric under the interchange of $\au_{a}^{i}$ and $\ad_{a}^{i}$.
This symmetry argument, and the fact that the Lagrange multiplier that should
appear in the constraint functional generating diffeomorphisms must be a vector
field, tells us that the generator of the diffeomorphisms on $\Sigma$ must have
the general structure:

\begin{equation}
D(\vec{N})=\int_{\Sigma}d^{3}x N^{a}\left\{\alpha\left[\ad_{a}^{i}\cd_{b}
\bu^{b}_{i}+\au_{a}^{i}\cu_{b}\bd^{b}_{i}\right]+\beta\left[\au_{a}^{i}\cd_{b}
\bu^{b}_{i}+\ad_{a}^{i}\cu_{b}\bd^{b}_{i}\right]\right\}
\label{15b}
\end{equation}

\noindent A simple computation gives $\alpha=1$, $\beta=0$. Notice that
$\cu_{b}\bd^{b}_{i}+\cd_{b}\bu^{b}_{i}=0$ and $\ad_{a}^{i}\cd_{b}\bu^{b}_{i}+
\au_{a}^{i}\cu_{b}\bd^{b}_{i}=0$ imply $\cd_{b}\bu^{b}_{i}=0$ and
$\cu_{b}\bd^{b}_{i}=0$ if $\det\left[\ad_{a}^{i}(x)-\au_{a}^{i}(x)\right]\neq
0$.

We check now that the algebra of $G(N)$ and $D(\vec{N})$ is the usual one; and
as a consequence of this we show that the constraints (\ref{10c}) are a first
class system. To this end we compute:

\begin{eqnarray}
& \left\{G(N), \au_{a}^{i}(x)\right\}=-\cu_{a}N^{i}\hspace{2cm}
& \left\{D(\vec{N}), \au_{a}^{i}(x)\right\}={\cal L}_{\vec{N}}\au_{a}^{i}(x)
\nonumber\\
& \left\{G(N), \ad_{a}^{i}(x)\right\}=-\cd_{a}N^{i}\hspace{2cm}
& \left\{D(\vec{N}), \ad_{a}^{i}(x)\right\}={\cal L}_{\vec{N}}\ad_{a}^{i}(x)
\label{16}
\end{eqnarray}

\noindent With the aid of (\ref{16}) it is straightforward to obtain:

\begin{eqnarray}
& & \left\{G(N),G(M)\right\}=-G([N,M])\nonumber\\
& & \left\{D(\vec{N}),G(M)\right\}=-G({\cal L}_{\vec{N}}N^{i})\label{17}\\
& & \left\{D(\vec{N}),D(\vec{M})\right\}=-D([\vec{N}, \vec{M}])\nonumber
\end{eqnarray}

\noindent Where $[N,M]^{i}=\epsilon^{i}_{\;\;jk}N^{j}M^{k}$ and
$[\vec{N}, \vec{M}]={\cal L}_{\vec{N}}\vec{M}$ is the commutator of the
vector fields $\vec{N}$ and $\vec{M}$.

\section{A Coordinate Transformation in The Phase Space}

In this section we introduce a convenient change of coordinates that will allow
us to pass from the usual connection-triad fields to the two-connection form of
the Husain-Kucha\v{r} model. This transformation will be used in the next
section to show that gravity itself can be described in the new phase space. We
will briefly discuss also an interesting type of canonical transformation that
naturally appears in this formalism.

We start by introducing the following functions in the $(\au_{a}^{i},
\ad_{a}^{i})$ phase space:

\begin{eqnarray}
& & {\rm A}_{a}^{i}(x)=\alpha(x)\au_{a}^{i}(x)+[1-\alpha(x)]\ad_{a}^{i}(x)=
\ad_{a}^{i}(x)+\alpha(x)[\au_{a}^{i}(x)-\ad_{a}^{i}(x)]\nonumber\\
& & \tilde{{\rm E}}^{a}_{i}(x)=\tilde{\eta}^{abc}\epsilon_{ijk}[\ad_{b}^{j}-
\au_{b}^{j}][\ad_{b}^{k}-\au_{b}^{k}]\label{18}
\end{eqnarray}

\noindent In the first equation we define a connection ${\rm A}_{a}^{i}$ as a
linear combination of two connections. This is possible because the
coefficients of $\au_{a}^{i}$ and $\ad_{a}^{i}$ satisfy the condition that
their sum is equal to one. The
equations (\ref{18}) define a change of coordinates only when the Jacobian:

\begin{equation}
\det  \left[ \begin{array}{cc}
\begin{picture}(40,34)(0,0)
\put(3,22){$\delta A_{\scriptscriptstyle{a}}^{i}(x)$}
\put(3,2){$\delta\au_{\scriptscriptstyle {b}}^{j}(y)$}
\put(0,17){\line(1,0){40}}
\end{picture} &
\begin{picture}(40,34)(0,0)
\put(3,22){$\delta A_{\scriptscriptstyle{a}}^{i}(x)$}
\put(3,2){$\delta\ad_{\scriptscriptstyle {b}}^{j}(y)$}
\put(0,17){\line(1,0){40}}
\end{picture}\\
\begin{picture}(40,34)(0,0)
\put(3,18){$\delta \tilde{E}^{\scriptscriptstyle{a}}_{i}(x)$}
\put(3,-2){$\delta\au_{\scriptscriptstyle {b}}^{j}(y)$}
\put(0,13){\line(1,0){40}}
\end{picture} &
\begin{picture}(40,34)(0,0)
\put(3,18){$\delta \tilde{E}^{\scriptscriptstyle{a}}_{i}(x)$}
\put(3,-2){$\delta\ad_{\scriptscriptstyle {b}}^{j}(y)$}
\put(0,13){\line(1,0){40}}
\end{picture}
\end{array} \right]
\label{18b}
\end{equation}

\noindent is different from zero. In our case (\ref{18b}) has the form:

\begin{equation}
\det  \left[ \begin{array}{cc}
\alpha I &
(1-\alpha) I\\
-\omega&
\omega
\end{array} \right]
\label{20}
\end{equation}

\noindent where $I\equiv\delta_{a}^{b}\delta_{i}^{j}\delta^{3}(x,y)$
and $\omega^{(a_{\scriptstyle{i}})(b_{\scriptstyle{j}})}=2\tilde{\eta}^{abc}
\epsilon_{ijk}\left(\ad_{c}^{k}(x)-\au_{c}^{k}(x)\right)\delta^{3}(x,y)$.
A straightforward
computation tells us that the previous Jacobian is equal to\footnote{we take
$\det I=1$} $\det\omega$. We see
then that the coordinate transformation introduced above is well defined if
and only if the symplectic structure $\Omega$ is non-degenerate.
It is straightforward to see that $\Omega$ can be written in terms of
${\rm A}_{a}^{i}$ and $\tilde{{\rm E}}_{i}^{a}$ as:

\begin{equation}
\Omega=\int_{\Sigma}d^{3}x\;d{\rm A}_{a}^{i}(x)\wedge d\tilde{{\rm
E}}^{a}_{i}(x)
\label{19}
\end{equation}

\noindent We notice now that the scalar field $\alpha(x)$ does
not appear in (\ref{19}). This means that a change in $\alpha(x)$ defines a
canonical transformation because it leaves the symplectic structure invariant.
It is also worthwhile pointing out that although (\ref{19}) seems to make
sense for
degenerate $\tilde{{\rm E}}_{i}^{a}$ ($\tilde{{\rm E}}_{i}^{a}$ is degenerate
when $\det\left[\ad_{a}^{i}(x)-\au_{a}^{i}(x)\right]=0$) it is only valid when
the change of coordinates introduced above is well defined.

With the help of (\ref{18}) we can write the Gauss law and the
diffeomorphism constraint in terms of $\au_{a}^{i}$ and $\ad_{a}^{i}$.
Substituting it in $\nabla_{a}\tilde{{\rm E}}^{a}_{i}=0$ we get
$\cu_{b}\bd^{b}_{i}+\cd_{b}\bu^{b}_{i}=0$ with no dependence on $\alpha(x)$.
This means that the transformation (equivalent to changing $\alpha(x)$ in
(\ref{18})):

\begin{eqnarray}
& & \au_{a}^{i}(x)\rightarrow \au_{a}^{i}(x)+\beta(x)\left[\au_{a}^{i}(x)-
\ad_{a}^{i}(x)\right]\nonumber\\
& & \ad_{a}^{i}(x)\rightarrow \ad_{a}^{i}(x)+\beta(x)\left[\au_{a}^{i}(x)-
\ad_{a}^{i}(x)\right]\label{22}
\end{eqnarray}

\noindent leaves the Gauss law invariant. The generator $P(\beta)$ of this
transformations must satisfy:

\begin{eqnarray}
& & \left\{P(\beta), \au_{a}^{i}(x)\right\}=\beta(x)\left[\au_{a}^{i}(x)-
\ad_{a}^{i}(x)\right]\nonumber\\
& & \left\{P(\beta), \ad_{a}^{i}(x)\right\}=\beta(x)\left[\au_{a}^{i}(x)-
\ad_{a}^{i}(x)\right]\label{22b}
\end{eqnarray}

\noindent We write $P(\beta)$ as:

\begin{equation}
P(\beta)=\int_{\Sigma}d^{3}x\;\beta(x)\Phi(\au, \ad)
\label{23}
\end{equation}

\noindent (we have ommited the indices and density weights that $\beta$
and $\Phi$ must carry in order to make the integrand a gauge invariant scalar
density of weight +1). The invariance of the Gauss law and the fact that the
infinitesimal parameter of the transformation $\beta$ is a scalar function
tells us that $\Phi$ carries no $SO(3)$ indices. If we take now the generator
of the diffeomorphisms $D(\vec{N})$ we see that under the action of (\ref{22})
it transforms as:

\begin{equation}
D(\vec{N})\rightarrow
D(\vec{N})-\frac{2}{3}\int_{\Sigma}d^{3}x(N^{a}\partial_{a}\beta)
\epsilon_{ijk}\tilde{\eta}^{bcd}[\ad_{b}^{i}-\au_{b}^{i}]
[\ad_{c}^{j}-\au_{c}^{j}][\ad_{d}^{k}-\au_{d}^{k}]
\label{24}
\end{equation}

\noindent Knowing that $\left\{D(\vec{N}),P(N)\right\}=-P\left({\cal
L}_{\vec{N}}N\right)$ for any functional $P$ we  can read directly from the
previous expression for $P(\beta)$\footnote{I am grateful to A. Ashtekar
and M. Varadarajan for discussions on this point}:

\begin{equation}
P(\beta)=\frac{2}{3}\int_{\Sigma}d^{3}x\beta(x)
\epsilon_{ijk}\tilde{\eta}^{bcd}[\ad_{b}^{i}-\au_{b}^{i}]
[\ad_{c}^{j}-\au_{c}^{j}][\ad_{d}^{k}-\au_{d}^{k}]
\label{25}
\end{equation}

\noindent As we can see $\Phi(\au, \ad)$ is a gauge invariant scalar density
of weight +1 in agreement to the argument presented above. It is
straightforward to check that $P(\beta)$ generates the infinitesimal gauge
transformations (\ref{22}).

In order to understand the origin of this symmetry
one can go to the action (\ref{1}) and perform the following transformation on
the 4-dimensional connection and frame fields:

\begin{eqnarray}
& & {\rm A}_{a}^{i}(x)\rightarrow  {\rm A}_{a}^{i}(x)+\epsilon \;e_{a}^{i}(x)
\nonumber\\
& & e_{a}^{i}(x)\rightarrow e_{a}^{i}(x)
\label{26}
\end{eqnarray}

\noindent ($\epsilon$ is an arbitrary scalar field).
Notice that the pull-back of these transformations onto the
3-dimensional slices $\Sigma$ is obtained from (\ref{18}) by varying $\alpha$.
Introducing (\ref{26}) in (\ref{1}) we see that the action transforms into:

\begin{equation}
\int_{{\cal M}}d^{4}x \tilde {\eta}^{abcd}\left[ {\rm F}_{ab}^{i}+2 \epsilon
\nabla
e_{b}^{i}+2(\partial_{a}\epsilon)e_{b}^{i}+\epsilon^{2}\epsilon^{i}_{\;\;lm}
e_{a}^{l}e_{b}^{m}\right]e_{c}^{j}e_{d}^{k}
\epsilon_{ijk}
\label{27}
\end{equation}

\noindent If $\epsilon$ is a constant then the third term in (\ref{27}) is
zero, the second one is a total divergence and the last one is identically
zero.
We thus see that in this case the
action is invariant under these transformations. Remember that both the Gauss
law and the diffeomorphism
constraints are invariant under (\ref{22}) when the parameter $\epsilon$ is a
constant. The changes in $\alpha$ in (\ref{18}) that
are independent of the point are global transformations; the conserved charge
associated with this symmetry is the volume of the "spatial" 3-manifold
(an obvious fact because there is no dynamics in the model).

\section{Two Connection Gravity}

In the previous two sections we have shown that the Husain-Kucha\v{r} model can
be interpreted in terms of two connections. We have seen also that there is a
natural way to translate results in the usual phase space $({\rm A}_{a}^{i},
\tilde{{\rm E}_{i}^{a}})$ to the $(\au_{a}^{i},\ad_{a}^{i})$ one. The main
result in this section is showing that gravity itself admits an interpretation
as a two connection theory. The idea is to work in the phase space of the
complexified Husain-Kucha\v{r} model and introducing the Hamiltonian constraint
"by hand" using equation (\ref{18}) to translate it to the two connection form.
A point that we want to discuss in this section is the role of the degenerate
metrics in the Ashtekar formulation of General Relativity. The possibility of
working with degenerate metrics is a feature that distinguishes Ashtekar's
connection dynamics from geometrodynamics. This may well be a welcome fact
because one could conceivably accommodate things such as topology changes and
evolution past some type of singularities in the formalism \cite{Ashlibro}. One
should point out, however, that degenerate metrics can also be a source of
trouble, for example when considering the issue of the existence  of the ground
state of the quantum theory \cite{Smo} because, as it has been shown by
Varadarajan \cite{Madh} there are classes of degenerate solutions to all the
constraints, in the spherically symmetric case, that are everywhere
non-singular but have arbitrary negative energy.

In our formulation it is possible to see that the non-degeneracy
condition for the symplectic form is equivalent to the condition that the
metric
is non-degenerate. The non-degeneracy condition for $\Omega$ is that at each
point of $\Sigma$ the 9$\times$9 matrix $\omega^{(a_{\scriptstyle{i}})
(b_{\scriptstyle{j}})}=2\;\tilde{\eta}^{abc}
\epsilon_{ijk}\left[\ad_{c}^{k}(x)-\au_{c}^{k}(x)\right]$ must be invertible.
Notice that, in principle, this is different from the non-degeneracy condition
for the $3\times3$ matrix ${\rm e}_{a}^{i}=\ad_{a}^{i}-\au_{a}^{i}$. We prove
now, however, that both conditions are equivalent\footnote{ This was
suggested to the author by A. Ashtekar.}. When $\det {\rm e}\neq 0$
the inverse of $\omega^{(a_{\scriptstyle{i}})(b_{\scriptstyle{j}})}$ is:

\begin{equation}
\omega^{-1}_{(a_{\scriptstyle{i}})(b_{\scriptstyle{j}})}=\frac{1}{4\det {\rm
e}}\left({\rm e}_{a}^{i}{\rm e}_{b}^{j}-2{\rm e}_{a}^{j}{\rm e}_{b}^{i}\right)
\label{30}
\end{equation}

\noindent We conclude then that the non-degeneracy of ${\rm e}_{a}^{i}$
implies the
non-degeneracy of $\omega_{(a_{\scriptstyle{i}})(b_{\scriptstyle{j}})}$. In
order to
prove the converse let us suppose that ${\rm e}_{a}^{i}$ is non-invertible and
different from zero (the ${\rm e}_{a}^{i}=0$ case is trivially dealt with)
then there exists an internal vector $v^{i}\neq 0$ such that ${\rm
e}_{a}^{i}v_{i}=0$. Let us consider now $v_{a}^{i}=\epsilon^{ijk}e_{aj}v_{k}$.
It is straightforward to show that if ${\rm e}_{a}^{i}\neq 0$ then
$v_{a}^{i}\neq 0$. Now:

\begin{eqnarray}
\omega^{(a_{\scriptstyle{i}})(b_{\scriptstyle{j}})}v_{a}^{i}& \!=\! &
2\tilde{\eta}^{abc}\epsilon_{ijk}\epsilon^{ilm}{\rm e}_{al}v_{m}{\rm
e}_{ck}=\nonumber\\
& \!=\!&2\tilde{\eta}^{abc}\left({\rm e}_{aj}v_{k}{\rm e}_{c}^{k}-{\rm
e}_{ak}{\rm e}_{c}^{k}v_{j}\right)=0\nonumber
\end{eqnarray}

\noindent because ${\rm e}_{a}^{i}v_{i}=0$ and the symmetry of ${\rm
e}_{ak}{\rm e}_{c}^{k}$ in a and c. We see that if
$\omega_{(a_{\scriptstyle{i}})(b_{\scriptstyle{j}})}$ is invertible then so is
${\rm e}_{a}^{i}$.
In the two connection phase space only non degenerate triads are allowed by the
non-degeneracy property of the symplectic 2-form.

We write now the Hamiltonian constraint in Ashtekar's description of General
Relativity in terms of $(\au, \ad)$ by using (\ref{18}):

\begin{eqnarray}
& & \eut_{abc}\epsilon^{ijk}\tilde{\rm E}^{a}_{i}\tilde{\rm E}^{b}_{j}
\tilde{\rm B}^{c}_{k}=8(\det {\rm e}){\rm e}_{c}^{k}\tilde{\rm B}^{c}_{k}=
\nonumber\\
& & 8(\det {\rm e}){\rm
e}_{c}^{k}\left[\alpha\bu^{c}_{k}+(1-\alpha)\bd^{c}_{k}+
\alpha(\alpha-1)\tilde{\eta}^{cde}\epsilon_{klm}{\rm e}_{d}^{l}{\rm e}_{e}^{m}
\right]\nonumber
\end{eqnarray}

\noindent where $\tilde{\rm B}_{i}^{a}=\tilde{\eta}^{abc}{\rm F}_{ab\;i}$. The
non
degeneracy of ${\rm e}_{a}^{i}$ allows us to write the Hamiltonian constraint
as: ${\rm e}_{a}^{i}\tilde{\rm B}^{a}_{i}=0$. We have the possibility of
choosing for $\alpha$ any value we want; for example $\alpha=1$. In this case
the Hamiltonian constraint reduces to ${\rm e}_{a}^{i}\bu^{a}_{i}=0$.
Summarizing, in the new phase space the constraints of General Relativity are:

\begin{eqnarray}
& & \cu_{\!a}\bd^{a}_{i}=0\nonumber\\
& & \cd_{\!a}\bu^{a}_{i}=0\label{31}\\
& & [\ad_{c}^{k}-\au_{c}^{k}]\bu_{k}^{c}=0\nonumber
\end{eqnarray}

\noindent Using the fact that: $$\cu_{\!a}\bd^{a}_{i}=\cu_{\!a}\bd^{a}_{i}-
\cd_{\!a}\bd^{a}_{i}=\epsilon_{ij}^{\;\;\;\;k}[\au_{a}^{j}-\ad_{a}^{j}]
\bu_{k}^{a}$$
(and the analogous expression for $\cd_{\!a}\bu^{a}_{i}$) the constraints
(\ref{31}) can be recast in the very neat form:

\begin{eqnarray}
& & \epsilon_{i}^{\;\;jk}e_{aj}\bu_{k}^{a}=0\nonumber\\
& & \epsilon_{i}^{\;\;jk}e_{aj}\bd_{k}^{a}=0\label{32}\\
& & e_{a}^{k}\bu_{k}^{a}=0\nonumber
\end{eqnarray}

\noindent As we can see the structure of the constraints is very simple. They
are either internal vector products of the curvatures and ${\rm e}_{a}^{i}$
or the internal scalar product of one of the curvatures and ${\rm e}_{a}^{i}$.
All the previous
expressions are densities of weight +1. At this point it is necessary to
stress
that although (\ref{32}) describes gravity in the two connection phase space we
do not have an action (as we had for the Husain-Kucha\v{r} model) that leads to
the previous Hamiltonian formulation of Gravity.

It is interesting to point out that we cannot make our mechanism work
if we take as the starting point the self-dual actions introduced
by Samuel, Jacobson, and Smolin \cite{Sam}, \cite{JaS}. As it has been
emphasized throughout the paper the key idea in the two-connection formulation
is that for $SO(3)$ the frame fields can be written as the difference of two
connections. In the self-dual action the fields have $SO(1,3)$ indices.
Although the symplectic structure can be still be written\footnote{
I, J are $SO(1,3)$ indices, $A^{IJ}_{a}$  is the self-dual connection and
$\tilde{E}_{IJ}^{a}=\tilde{\eta}^{abc}e_{bI}e_{cJ}$} as:

\begin{equation}
\Omega=2\int_{\Sigma}d^{3}x\;\tilde{\eta}^{abc}e_{bI}de_{cJ}\wedge d
A^{IJ}_{a}=\int_{\Sigma}d^{3}x\;d\tilde{E}_{IJ}^{a}\wedge d
A^{IJ}_{a}
\nonumber
\end{equation}

\noindent we cannot use $(A^{IJ}_{a}, e_{bK})$ as the phase space (it is not
even-dimensional!)  and it is not straightforward to relate the
non-degeneracy of the symplectic structure with the non-degeneracy of
$\tilde{E}_{IJ}^{a}$.

The new constraints (\ref{32}) {\it look} very simple, but, are they
really so? In the case of the Ashtekar constraints the variables that are used
are canonically conjugate so that when quantizing the theory, for example, in
the connection representation, the quantum operators that describe the
constraints are very simple. In our case it is not clear what set of
elementary variables makes the quantization of the theory easy. Only if a
suitable set of elementary variables and a representation of them as operators
acting in a vector space can be found such that the quantum constraints are
simple should we say that a simplification has occured because of
the introduction of (\ref{32}). Our hope is that the availability of geometric
objects in the two-connection phase space that are not obvious in the usual
Ashtekar phase space will lead to sets of elementary variables that would
allow us to advance in the quantization of gravity.

\section{Conclusions}

We have shown that it is possible to describe gravity (and some other
diff-invariant theories in 3+1 dimensions like the Husain-Kucha\v{r} model) in
a phase space spanned by two different $SO(3)$ connections. Due to the form
of the symplectic structure $\Omega$ , non-degenerate frame fields are excluded
by the non degeneracy property of $\Omega$. This allows us to simplify
the Hamiltonian constraint and write it in a form that is very close to the
one of the remaining constraints. The  relevance of these results relies mainly
on the fact that having a new phase space it is conceivable that new systems of
elementary variables can be found that will allow us to attack the quantization
of gravity from a different perspective. The hope is that, in analogy with what
happened with the introduction of the Ashtekar variables and the subsequent
introduction of the loop representation (that so many new results concerning
the structure of the space-time etc... have given to us) this new phase space
description will help in gaining new information about Quantum Gravity.

\newpage

{\bf Acknowledgements}
I am indebted to Abhay Ashtekar for many illuminating remarks and
suggestions. I would also like to thank Madhavan Varadarajan for the
many interesting discussions we have had and
V. Husain, G. Immirzi, J. Louko, N Manojlovi\'c,  G. Mena, H. Morales,
L. Smolin, and R. Sorkin for several useful remarks and comments.
I am grateful to the Spanish
Research Council (C.S.I.C.) for providing financial support. This work
was done between Syracuse University and Penn State University. I am
grateful to them for their hospitality.

\end{document}